\documentclass[english]{article}
\usepackage[]{fontenc}
\usepackage[latin1]{inputenc}
\usepackage{babel}
\usepackage{epsfig}

\def \bi{\bibitem}
\def \Tr{{\rm Tr}}
\def \l{\lambda}
\def \C{{\bf C}}


\begin{document}

\title{Ultrametricity Between States at Different Temperatures in Spin-Glasses}
\author{
Tommaso Rizzo\\
\small Dipartimento
di Scienze Fisiche, Universit\`a ``Federico II'',\\
\small Complesso Monte S. Angelo, I--80126 Napoli (Italy)\\
\small e-mail: {\tt tommaso.rizzo@inwind.it}}

\maketitle

\begin{abstract}
We prove the existence of correlations between the equilibrium states at different temperatures
of the multi-$p$-spin spherical spin-glass models with continuous replica
symmetry breaking: there is no chaos in temperature in these models. Furthermore, the overlaps satisfy ultrametric relations. 
As a consequence the Parisi tree is essentially the same at all temperatures with lower branches developing when lowering the temperature.  
We conjecture that the reference free energies of the clusters are also fixed at all temperatures as in the generalized random-energy model. 
\end{abstract}

\section*{Introduction}

The effects of temperature changes currently are the most interesting in spin-glass studies. 
The coexistence of chaos (or rejuvenation) and memory \cite{LSB83,MEeCH,Misc1} is particularly puzzling.
A first non-trivial problem is whether these effects are universal, i.e. whether they are present in different real spin-glasses \cite{VIN} and in numerical simulations of microscopic models \cite{FelFed}; 
the main problem, however, lies in the theoretical explanation of these phenomena. 
If one admits that the static (equilibrium) free-energy landscape is somehow relevant to the off-equilibrium dynamics, the two effects of chaos and memory seem to contradict each other: the
former points in the direction of absence of correlation between the
equilibrium states at different temperature, while the latter points in the opposite
direction. 
A qualitative explanation of the coexistence of the two effects has been given in terms of a hierarchical picture in which states at different temperatures are organized on the same ultrametric tree, whose details are revealed by lowering the temperature \cite{BOU, Met, MEeCH}.
Recent numerical and theoretical work on the subject has
confirmed that the generalized random energy model (GREM) \cite{GREM1,GREM2}, a model with such a landscape structure, displays chaos as well as memory \cite{grem}. 
However, the GREM is defined through its phase space, without any reference
to any underlying microscopic description; therefore it would be interesting to find out whether there are spin-glasses that behave like the GREM.
The question is whether equilibrium states at different temperatures are correlated or not and which is the nature of the correlations. The hypothesis of chaos in temperature affirms that they are uncorrelated in finite dimensional systems \cite{BM87,FH}; this has been studied analytically in a number of paper \cite{Kondor1,Kondor2,FN}; We have questioned the hypothesis validity in the Sherrington-Kirkpatrick (SK) model in a recent paper \cite{mio1}, henceforth labeled [I]. The problem has also been investigated numerically \cite{MarBill,Rit,Ney}.  
It has also been shown that the random energy model (REM), which is non-chaotic by definition, is capable of producing strong rejuvenation signals \cite{BouSal}.
On the other hand various attempts are underway to substantiate the idea of some kind of hierarchical structure underlying the phenomenology in the context of real-space theories like the droplet model \cite{Misc3,temasmic}.

In this paper we consider the mean-field multi-\( p \)-spin
spherical spin-glass models for values
of the coupling constants such that they display full replica symmetry breaking (RSB) \cite{theo}.
It must be remarked that the analysis of the off-equilibrium dynamics \cite{CKBM} of these models has revealed the coexistence of chaos and memory \cite{CK}.
We study the correlations between
the equilibrium states at different temperatures following the lines developed in {[}I{]}, where the problem
has been put in connection with the existence of a particular class of solutions
of the saddle point (SP) equations obtained in the replica framework.
In section 1 we prove that these solutions exist for the class
of models considered, implying strong correlations between states at different temperatures, i.e. no chaos in temperature. 

These solutions are built with Parisi matrices \cite{MPV}
and in section 2 we show that this structure implies ultrametric relations between the overlaps of states
at different temperatures. 
These ultrametric relations determine a one-to-one correspondence (except at the lower levels) between the trees of states at two different temperatures in such a way that to a cluster of states at temperature
\( T_{1} \), at a level of the $T_1$ tree labeled by some value \( q_{1} \) of the overlap, corresponds
a cluster of states at temperature \( T_{2} \) at  a level of the $T_2$ tree labeled by the same
value \( q_{2}=q_{1} \). We call {}``twins{}" any two clusters in such a correspondence. 
Specifically, a state at temperature \( T_{1} \) corresponds to a
cluster of states at temperature \( T_{2}<T_{1} \) whose minimal
overlap is given by \( q_{EA1} \). Therefore each state at temperature
\( T_{1} \) can be considered as the father of many states at temperature
\( T_{2} \). A son has only one father, and it has an overlap lower
than \( q_{EA1} \) with all the other states at \( T_{1} \). 
The
meaning of the correspondence between the trees at different temperatures is that any
relationship between a state or cluster \( \alpha  \) at \( T_{2} \) and any other state or cluster 
\( \gamma  \) at \( T_{1} \) or $T_2$ is univocally determined by the relationship
between \( \gamma  \) and the twin of \( \alpha  \) at $T_1$. In particular,
given a state \( \gamma  \) at $T_2$ and its father \( \Gamma  \) at $T_1>T_2$, the overlap
between \( \gamma  \) and any state \( \Lambda  \) at
\( T_{1} \) is given by \( q_{\Lambda \gamma }=q_{\Lambda \Gamma } \). 

Each state of a spin-glass
system at a given temperature has a certain statistical weight \( P_{\alpha } \) determined by the Gibbs measure.
According to the Parisi solution \cite{MPV} the fluctuations of the weights
with the disorder can be described through a stochastic process involving
the free energies \( f_{\alpha } \) of the states, which are defined
as \( P_{\alpha }=\exp [-\beta f_{\alpha }]/\sum _{k}\exp [-\beta f_{k}] \) \cite{fren1,fren2}.
Given the tree of states, one considers a given level $k$ and assigns
independently to each cluster of states at this level a reference free energy chosen
randomly such that the average number of clusters with free energy
between \( f \) and \( f+df \) is given by
\begin{equation}
d{\mathcal{N}}(f)=\exp [\beta x(q_{k})f]df
\label{legge1}
\end{equation}
Where $x(q_k)$ is the inverse of the Parisi function $q(x)$.
After having applied the procedure to every level of the tree,
the free energy of a given state is set to be the sum of the reference free
energies of the clusters to which it belongs at the various levels.
For instance, if the state \( \alpha  \) belongs to the cluster \( i_{1} \)
at the first level of the tree, to the cluster \( i_{1}i_{2} \) at
the second level of the tree and so on, its free energy is given by
\begin{equation}
f_{\alpha }=f_{i_{1}}+f_{i_{1}i_{2}}+f_{i_{1}i_{2}i_{3}}+\ldots +f_{i_{1}i_{2}i_{3}\ldots i_{L-1}}+f_{i_{1}i_{2}i_{3}\ldots i_{L-1}\alpha }.
\end{equation}
The distribution obtained has various interesting features; for instance, in each sample there are only few states with a finite weight, while there is an infinite number of states that carries an infinitesimal weight.
 
Since we found a one-to-one correspondence between the trees of states at different temperatures, it is natural to whether if there are also correlation between the weights. 
In the replica framework the standard procedure to cope with this
problem requires the computation of sums over all the different
solutions of the saddle point equations, in order to obtain the cumulants
of the distribution function of the weights. When considering a system at a given temperature,
this can be done noticing that all the solutions are permutations
of the standard Parisi solution \cite{YD,MPV}; therefore the sum over solutions can be replaced by a sum over replica indices; Unfortunately when considering
systems at different temperatures the solutions are not permutations
one of the other and we are unable to sum them all. However, in section 3 we conjecture
that the weights distribution is identical to that of the GREM
class of models.
In the GREM the tree of states is the same at all temperatures by definition and the reference free energies of the clusters are constant too \cite{GREM1,GREM2}. 
Their distribution obeys the law (\ref{legge1}) with $\beta x(q)$ replaced by $y_u(q)$, a function that depends on the model.
In analogy to this we guess that in the models we are considering the reference free energy of a given cluster at $T_1$ is equal to the reference free energy of its twin at $T_2$; we notice that since for this models we have $\beta x(q,T)=y_u(q)$, (where $y_u(q)$ depends on the coupling constants but not on the temperature), this is consistent with the fact that the same set of free energies obeys law (\ref{legge1}) at all temperatures. 
This belief is motivated by the fact that there is
a one-to-one correspondence between the solutions of SP equations of the 
multi-\( p \) spin spherical model and the solutions one finds in
the analogous replica treatment of the GREM \cite{FN}.
Indeed, the solutions have the same {\em formal} structure expressed
in terms of the function \( y_{u}(q) \) and the differences between the models  shows up only in the actual values of this function.
Since for every solution of the multi-$p$ spherical spin-glass models there is a correspondent solution of the GREM, we expect that the sums over solutions give equal results in the two models; more precisely we expect the results to have the same dependence on the function $y_u(q)$. Accordingly, the correlations between the weights  at different temperatures will be equal in the two classes of models. 

The previous conjecture could be proved if we were able to carry out the sum over all the solutions. This problem is common to many models where one would like to recover within the replica framework
a result which is known {\em a priori} (e.g. in the REM \cite{REM}), or which can be obtained through different methods (e.g. the spherical $p$-spin model where the absence of chaos in easily seen within the TAP approach \cite{MPK}). 
In section 4 we discuss some technical difficulties connected with this problem.
At the end we give our conclusions.
Technical details are skipped to the appendices.

\section{Correlations between states at different temperatures}

We recall the method discussed in {[}I{]} to study the correlations
between states at different temperatures. The
replica trick is usually used in order to compute the average over the disorder of the free
energy by computing the moments of the partition function
\( \overline{Z^{n}} \). In the thermodynamic limit saddle
point (SP) equations are obtained for the order parameter which is an \( n\times n \)
matrix \( Q_{ab} \). An outcome of the computation is that the order
parameter is connected to the distribution of the overlaps through
the relation 

\begin{equation}
\label{pq1}
q^{(k)}=\int q^{k}P(q)dq=\lim _{n\rightarrow 0}\sum_{\textrm{all solutions}}Q_{ab}^{k}
\end{equation}
Where the function \( P(q) \) is the averaged probability density of finding
two states with overlap \( q \) according to their Gibbs weight, and
\( a \) and \( b \) are two different replica indices (e.g. 1 and
2). In the r.h.s. of (\ref{pq1}) there appears a sum over all the different solutions of the saddle point equation with
the same (maximum) free energy; indeed, as soon as replica symmetry is broken,
we have many solutions: given a certain solution
others can be obtained through a permutation of the replica indices.
However, in the isothermal case, all the solutions are given by the
Parisi solution plus its natural permutations; therefore the sum over
them can be replaced with an average over the replica indices multiplied by the total number of different solutions, which in the $n\rightarrow 0$ limit goes to $1$ \cite{YD,MPV}. In {[}I{]}
it has been shown that the correlations between states at different
temperatures can be investigated through the computation of the \( n\rightarrow 0 \)
limit of the quantity \( (1-\overline{Z^{n}(T_{1})Z^{n}(T_{2})})/n \).
As a matter of fact, we are not interested in its actual value (which must be equal
to the sum of the free energies at temperature \( T_{1} \) and \( T_{2} \)), but in the saddle point equation obtained by applying the replica trick
to the whole \( \overline{Z^{n}(T_{1})Z^{n}(T_{2})} \) and not to
\( \overline{Z^{n}(T_{1})} \) and \( \overline{Z^{n}(T_{2})} \)
separately\footnote{
This quantity is actually the logarithm of the partition function of the two systems. To make the text more readable we will refer to it as the ``free energy'' of the two systems, while the free energy of a single system is the logarithm of its partition function {\em multiplied} by $\beta$, the inverse of its temperature. Therefore a claim like ``the free energy of the two systems must be equal to the sum of the free energies at temperature $T_1$ and $T_2$'' is shorthand for ``the logarithm of the partition function of the two systems must be equal to the sum of the logarithms of the partition functions at temperature $T_1$ and $T_2$''.}. 
In this case one obtains SP equations for a \( 2n\times 2n \)
matrix \( \hat{Q}=\left( \begin{array}{cc}
Q_{1} & P\\
P^{t} & Q_{2}
\end{array}\right)  \), where \( Q_{1}, \) \( Q_{2} \) and \( P \) are \( n\times n \)
matrices. The advantage of this procedure is that the matrix \( P \)
encodes information on the correlations between states at different
temperatures; indeed, through the same steps that led to (\ref{pq1}) it can be shown that \[
q_{T1T2}^{(k)}=\int q_{T1T2}^{k}P(q_{T1T2})dq_{T1T2}=\lim _{n\rightarrow 0}\sum _{\textrm{all solutions}}P_{ab}^{k}	\]
Where the function \( P(q_{T1T2}) \) is the generalization of the
\( P(q) \) to states at different temperatures. As above, 
we have to sum  over all the solutions of the SP equations:
this is quite a delicate point; indeed we do not expect that all solutions
be natural permutations of a single solution as in the isothermal case (i.e. the standard Parisi solution).
Actually, there are infinite solutions corresponding to different parameterizations
and we don't know how to sum them all. However, it is possible
to reconstruct the function \( P(q_{T1T2}) \) in an indirect way
by knowing the weight distribution of the equilibrium states.
We have applied the method to the spherical spin-glass model with multi-\( p \)-spin
interaction \cite{theo}. The Hamiltonian of the model is defined
as\[
H=\sum _{p=2}^{\infty }\sum _{i_{1}<i_{2}<\ldots <i_{p}}J_{i_{1}i_{2}\ldots i_{p}}S_{i_{1}}S_{i_{2}}\ldots S_{i_{p}}+h\sum _{i}S_{i}\]
The \( J \)'s are independent Gaussian random variable with zero
mean and variance \( \left\langle J_{i_{1}i_{2}\ldots i_{p}}^{2}\right\rangle =(p-1)!J_{p}^{2}N^{1-p} \).
The spins are subjected to the spherical constraint \( \sum _{i}S_{i}^{2}=N\sigma  \).
Introducing the function \[
f(q)=\sum ^{\infty }_{p=2}\frac{1}{p}J_{p}^{2}q^{p}\]
The replicated free energy reads \[
2\beta F_{n}=-\beta ^{2}\sum _{ab}[f(Q_{ab})+H^{2}Q_{ab}]-\Tr\ln Q-n\]
where the variational parameter is an \( n\times n \) matrix \( Q_{ab} \)
with diagonal \( q_{d}=\sigma  \). The SP equations then read\begin{equation}
\label{spsingle}
\beta ^{2}f'(Q_{ab})+\beta ^{2}H^{2}=-\left( \frac{1}{Q}\right) _{ab}
\end{equation}
Expressing \( Q_{ab} \) as a Parisi function we obtain the solution
as \( q(x,T)=q_{u}(\beta x) \) where \( q_{u}(y) \) is defined as
the inverse of the universal function\[
y_{u}(q)=\frac{f'''(q)}{2f''(q)^{3/2}}\]
with the value of \( q(1) \) fixed by\[
q_{d}-q(1)=\frac{1}{\beta \sqrt{f''(q(1))}}\]
In the presence of a small magnetic field there is also a small plateau of temperature-independent height \( q_{H} \) fixed by the condition\[
H^{2}=q_{H}f''(q_{H})-f'(q_{H})\]
When we consider simultaneously two systems at different temperature,
we get the following expression for \( \overline{Z^{n}(T_{1})Z^{n}(T_{2})} \)
in the thermodynamic limit \[
F_{n}=\ln \overline{Z^{n}(T_{1})Z^{n}(T_{2})}=-\beta _{1}^{2}\sum _{ab}(f(Q_{1ab})+H^{2}Q_{1ab})-\beta _{2}^{2}\sum _{ab}(f(Q_{2ab})+H^{2}Q_{2ab})+\]
\begin{equation}
-2\beta _{1}\beta _{2}\sum _{ab}(f(P_{ab})+H^{2}P_{ab})-\Tr\ln \hat{Q}-n
\label{FUN2}
\end{equation}
The corresponding SP equations are similar to those for the isothermal
case (\ref{spsingle}) and are reported in appendix 2. A solution
of the SP equations is certainly the one with \( P=0 \) and \( Q_{1} \)
and \( Q_{2} \) equal to the corresponding isothermal solutions.
Its free energy is given by the sum of the free energies at temperatures
\( T_{1} \) and \( T_{2} \) as expected. The problem is whether
other solutions exist with a non-zero \( P \) and with the same free
energy of the \( P=0 \) solution. In {[}I{]} a particular
structure was proposed for these \( P\neq 0 \) solutions (called non-chaotic, since
their existence implies absence of chaos in temperature); we
found that solutions having such a structure 
do exist for this class of models. Actually there is an infinite number of  solutions which are parameterized
by the value \( p_{d} \) of the diagonal of \( P \), whose value
ranges from zero to a maximum one, which, for these models, turns out
to be the self-overlap of the states at the higher temperature. The solution with $P=0$ is included in this set and corresponds to the value $p_d=0$. It
is very interesting to notice that they exist also when the two systems
at different temperatures are subjected to the same magnetic field
(see appendix 2). Here for simplicity we will refer to the \( H=0 \) case.
For a given value of \( p_{d} \) the solutions in terms of the three
functions \( q_{1}(x) \), \( q_{2}(x) \) and \( p(x) \) are\begin{eqnarray}
q_{s}(x) & = & \left\{ \begin{array}{ccc}
q_{u}((\beta _{1}+\beta _{2})x) & \textrm{for} & x\leq \frac{1}{\beta _{1}+\beta _{2}}y_{u}(p_{d})\\
p_{d} & \textrm{for} & \frac{1}{\beta _{1}+\beta _{2}}y_{u}(p_{d})\leq x\leq \frac{1}{\beta _{s}}y_{u}(p_{d})\\
q_{u}(\beta _{s}x) & \textrm{for} & \frac{1}{\beta _{s}}y_{u}(p_{d})\leq x\leq x_{max}(T_{s})
\end{array}\right. 
\label{solq}
\\
p(x) & = & \left\{ \begin{array}{ccc}
q_{u}((\beta _{1}+\beta _{2})x) & \textrm{for} & x\leq \frac{1}{\beta _{1}+\beta _{2}}y_{u}(p_{d})\\
p_{d} & \textrm{for} & \frac{1}{\beta _{1}+\beta _{2}}y_{u}(p_{d})\leq x\leq 1
\end{array}\right. \label{solp}
\end{eqnarray}
 \begin{figure}[htb]
\begin{center}
\epsfig{file=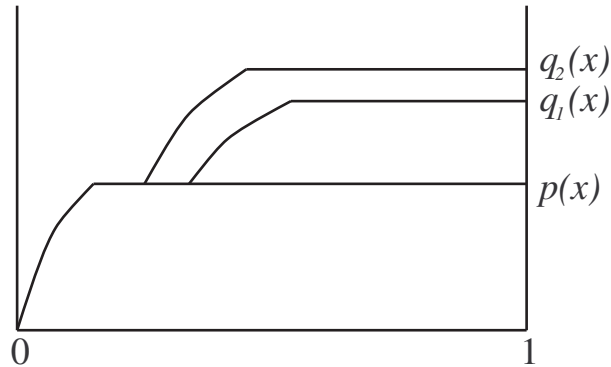}
\caption{The solutions for $p_d\neq 0$. In the small-$x$
region they are all equal to $q_u((\beta_1+\beta_2)x)$ until the point $x_c$ where
$p(x)=p_d$, for $x>x_c$ $p(x)$ is constant and equal to $p_d$, while $q_1(x)$ and
$q_2(x)$ after an intermediate plateau are joined respectively to $q_u(\beta_1x)$ and $q_u(\beta_2x)$; $p_d$ can take values between
zero and $q_{1EA}$.} \label{figure}
\end{center}\end{figure}
Where \( q_{u}(y) \) and \( y_{u}(q) \) are the universal function
defined above. These solutions can be built for any couple of temperatures
\( T_{1}\geq T_{2} \) both below the critical temperature and for
any \( p_{d} \) between zero and \( q_{1}(1) \) which is the self-overlap
of the states of the system at the higher temperature. The solutions
are sketched in figures 1 and 2. Notice that $q_1(x)$,$q_2(x)$ and $p(x)$ are
all equal in the small-\( x \) region. This is connected with the
fact that this class of models verifies the scaling \( q(x,T)=q(x/T) \). We skip
to appendix 2 the demonstration that these functions solve the SP
equations and have the correct free energy. 
It must be remarked that the solutions
 (\ref{solq},\ref{solp}) are formally the same which have been found with a similar
treatment of the GREM \cite{FN}, with a different model-dependent function $y_u(q)$.
 At
this stage we can infer that the function \( P(q_{T1T2}) \) has a
non-zero support from zero to the self-overlap of the states at the
higher temperature.

\section{Ultrametricity}

The solutions we found imply ultrametricity between states at
different temperatures of a given system. To see this we must consider the probability \( P(q_{12},q_{13},q_{23}) \)
of extracting three states at different temperatures with assigned values
of their mutual overlap. If we assume, for instance, that states
\( 1 \) and \( 2 \) are at temperature \( T_{1} \) and state
\( 3 \) is at temperature \( T_{2} \), this function is related
to the solutions of the SP equation through the following relation\[
\int q_{12}^{r}q_{13}^{s}q_{23}^{t}P(q_{12},q_{13},q_{23})dq_{12}dq_{13}dq_{23}=\lim _{n\rightarrow 0}\sum_{\textrm{all solutions}}Q_{1ab}^{r}P_{ac}^{s}P_{cb}^{t}\]
We are not able to perform the sum over solutions in r.h.s of
the previous expression, but we can infer ultrametricity by simply
looking at their structure. The function \( P(q_{12},q_{13},q_{23}) \)
can be reconstructed later in an indirect way from the distribution of the
weights we will describe below (section 3). 

\begin{figure}[bt]
\begin{center}
\begin{picture}(260,260)(0,0)

\put(0,0){\dashbox{120}(120,120)}
\put(0,140){\dashbox{120}(120,120)}
\put(140,0){\dashbox{120}(120,120)}
\put(140,140){\dashbox{120}(120,120)}

\multiput(0,60)(60,-60){2}{\dashbox{60}(60,60)}
\multiput(0,200)(60,-60){2}{\dashbox{60}(60,60)}
\multiput(140,60)(60,-60){2}{\dashbox{60}(60,60)}
\multiput(140,200)(60,-60){2}{\dashbox{60}(60,60)}

\multiput(0,90)(30,-30){4}{\dashbox{30}(30,30)}
\multiput(0,230)(30,-30){4}{\dashbox{30}(30,30)}
\multiput(140,90)(30,-30){4}{\dashbox{30}(30,30)}
\multiput(140,230)(30,-30){4}{\dashbox{30}(30,30)}

\multiput(0,245)(15,-15){8}{\dashbox{15}(15,15)}
\put(0,260){\line(1,-1){120}}
\multiput(0,245)(15,-15){8}{\line(1,1){15}}

\multiput(140,110)(10,-10){12}{\dashbox{10}(10,10)}
\put(140,120){\line(1,-1){120}}
\multiput(140,110)(10,-10){12}{\line(1,1){10}}

\put(27,27){$q(0)$}
\put(87,87){$q(0)$}
\put(12,72){$q_i$}
\put(42,102){$q_i$}
\put(72,12){$q_i$}
\put(102,42){$q_i$}
\multiput(12,102)(30,-30){4}{$p_d$}
\put(-10,-3){$n$}
\put(-12,57){$m_i$}
\put(-12,87){$m_c$}

\put(27,167){$q(0)$}
\put(87,227){$q(0)$}
\put(12,212){$q_{i}$}
\put(42,242){$q_{i}$}
\put(72,152){$q_{i}$}
\put(102,182){$q_{i}$}
\put(-10,137){$n$}
\put(-12,197){$m_i$}
\put(-12,227){$m_c$}

\put(167,27){$q(0)$}
\put(227,87){$q(0)$}
\put(152,72){$q_{i}$}
\put(182,102){$q_{i}$}
\put(212,12){$q_{i}$}
\put(242,42){$q_{i}$}
\put(130,-3){$n$}
\put(128,57){$m_i$}
\put(128,87){$m_c$}

\put(167,167){$q(0)$}
\put(227,227){$q(0)$}
\put(152,212){$q_i$}
\put(182,242){$q_i$}
\put(212,152){$q_i$}
\put(242,182){$q_i$}
\multiput(152,242)(30,-30){4}{$p_d$}
\put(130,137){$n$}
\put(128,197){$m_i$}
\put(128,227){$m_c$}

\end{picture}
\caption{The global matrix $\hat{Q}$ for a given $p_d$. $Q_1$ is at the left high corner, $Q_2$ at the low right corner, $P$ at the left low corner. If we consider two replicas $a$ and $b$ at temperature $T_1$ such that $Q_{1ab}\geq p_d$ the overlaps $P_{ac}$ and $P_{cb}$ with a replica $c$ at temperature $T_2$ verify $P_{ac}=P_{bc}\leq p_d$; if $Q_{1ab}\leq p_d$ then we may have $P_{ca}=Q_{1ab}\leq P_{cb}$, or $P_{cb}=Q_{1ab}\leq P_{ca}$, or $P_{cb}=P_{ca}\leq Q_{1ab}$; i.e. we have ultrametric relations between replicas at different temperatures.}\label{figure4}
\end{center}
\end{figure}
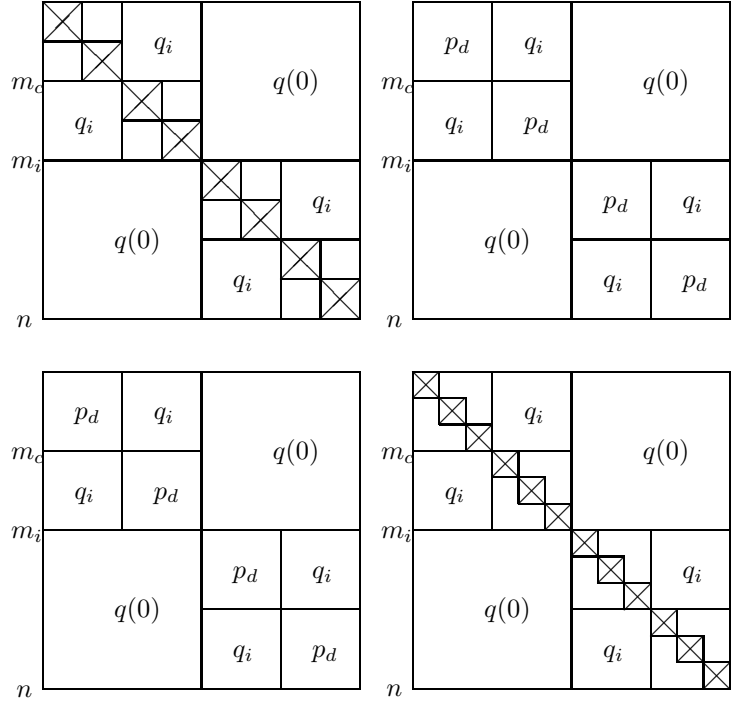

Let us consider the solution for a given
\( p_{d} \): a simple analysis of figure (2) shows that for any three
replica indices \( a \), \( b \), and \( c \), the corresponding
overlaps \( Q_{1ab} \),\( P_{ac} \) and \( P_{cb} \) always form
an isosceles triangle with the two equal sides smaller or equal to the third.
However, while \( Q_{1ab} \) can take values from zero to \( q_{EA1} \),
\( P_{ac} \) and \( P_{cb} \) can take values from zero to \( p_{d} \). Considering other solutions, we obtain values of $P$ between $0$ and $q_{1EA}$.
 Now, since ultrametricity is a property of all the solutions, it will
be a property of the sum over solutions as well; therefore the overlaps between any
three states at any temperature satisfy ultrametric relations.
In general, the maximum overlap between a state at temperature \( T_{1} \)
and another state at temperature \( T_{2} \) is equal to the self-overlap of the state at the higher
temperature.  Perfect ultrametricity between states at different temperatures is a special property of this
class of models which relies on the identity of the three function
\( q_{1}(x) \), \( q_{2}(x) \) and \( p(x) \) in the small-\( x \)
region. 

As we said in the introduction, the ultrametric relations between the overlaps of states at different temperatures define a correspondence between the trees of states.
Consider a cluster $I_1$ of equilibrium states at temperature $T_1$ whose overlaps are greater or equal to some $q_1$. Given a state $\alpha$ in $I_1$, we have that the overlap between any  other state $\beta$ in $I_1$ and any state $\gamma$ outside $I_1$ is simply given by $q_{\beta \gamma}=q_{\alpha \gamma}$ \cite{MPV}. Now, given a state $\alpha'$ at $T_2$ whose overlap with any of the states in $I_1$ is greater or equal to $q_1$, we have that the overlap between $\alpha'$ and any state $\gamma$ at $T_1$ outside $I_1$ is simply given by $q_{\alpha'\gamma}=q_{\alpha \gamma}$; In other words, from a geometrical point of view we may assume that $\alpha'$ {\em is} in $I_1$. Furthermore, this is true for all the states at $T_2$ whose overlap with $\alpha'$ is greater or equal than $q_1$; by definition they form a cluster $I_2$ which we call the {}``twin{}"  of $I_1$. As far as the overlaps are concerned the two clusters $I_1$ and $I_2$ can be considered the same; therefore from a topological point of view there is only one tree of states whose details are revealed by lowering the temperature.   
Finally, we recall that the solutions we described satisfy the separability property \cite{SEP,mio1}, which means that the overlap contains all the information between two states at equal or  different temperatures. 

\section{The distribution of the weights}

Each state of a spin-glass system at a given temperature has a certain
statistical weight \( P_{\alpha } \). The fluctuation of the weights
with the disorder can be described through the following procedure \cite{MPV,fren1,fren2}
involving the free energies of the states defined as \( P_{\alpha }=\exp [-\beta f_{\alpha }]/\sum _{k}\exp [-\beta f_{k}] \).
Given the tree of states, we consider a given level and assign to
each cluster of states at this level a reference free energy chosen
randomly such that the average number of clusters reference with free energy
between \( f \) and \( f+df \) is given by

\[
d{\mathcal{N}}(f)=\exp [\beta x(q_{k})f]df\]
After applying the procedure to any level of the tree we define
the free energy of a given state as the sum of the reference free
energies of the cluster to which it belongs at the various levels.
For instance, if the state \( \alpha  \) belongs to the cluster \( i_{1} \)
at the first level of the tree, to the cluster \( i_{1}i_{2} \) at
the second level of the tree and so on, its free energy is given by
\begin{equation}
f_{\alpha }=f_{i_{1}}+f_{i_{1}i_{2}}+f_{i_{1}i_{2}i_{3}}+\ldots +f_{i_{1}i_{2}i_{3}\ldots i_{L-1}}+f_{i_{1}i_{2}i_{3}\ldots i_{L-1}\alpha }
\end{equation}
We have established that the trees of states at different temperatures
are equal: to a cluster of states at \( T_{1} \) corresponds a cluster
of states at \( T_{2} \); therefore we want to know which is the
relation between the corresponding reference free energies. This correlation
can be obtained computing the quantities \[
M_{kj}=\overline{\sum _{I}W_{I,1}^{k}W_{I,2}^{j}}\]
 Where the index \( I \) refers to the clusters at a certain level
\( q \) of the tree and \( W_{I,1} \) and \( W_{I,2} \) are the
weight of the cluster \( I \) respectively at temperature \( T_{1} \)
and \( T_{2} \). Through standard manipulation \cite{MPV} we obtain
\begin{eqnarray*}
M_{kj} & = & \overline{\sum _{a_{1}\ldots a_{k}b_{1}\ldots b_{j}}P_{a_{1}}\ldots P_{a_{k}}P_{b_{1}}\ldots P_{b_{j}}\Theta (q_{a_{1}a_{2}}-q)\ldots \Theta (q_{a_{1}a_{k}}-q)\Theta (q_{a_{1}b_{1}}-q)\ldots \Theta (q_{a_{1}b_{j}}-q)}=
\\
 & = & \lim _{n\rightarrow 0}\sum_{\textrm{all solutions}}
\Theta (Q_{1a_{1}a_{2}}-q)\ldots \Theta (Q_{1a_{1}a_{k}}-q)\Theta (P_{a_{1}b_{1}}-q)\ldots \Theta (P_{a_{1}b_{j}}-q)
\end{eqnarray*}
Again, the problem is reduced to the computation of a sum
over all the solutions of the SP equations. We are unable to perform such
a sum but we have a line of reasoning to guess which may be the result.
Indeed, the previous expression does not depend on the model under consideration;
it is a general outcome of the replica trick. Therefore, if we have
two different models with the same set of solutions the corresponding weight
distribution functions will be equal. The solutions
 (\ref{solq},\ref{solp}) found above are formally the same which have been found with a similar
treatment of the GREM \cite{FN}, with a scaling \( q_{1}(x)=q_{2}(x)=p(x)=q_{u}((\beta _{1}+\beta _{2})x) \)
in the small-\( x \) region. In other words, for any solution with
\( P\neq 0 \) of the spherical model with multi-spin interactions
there is a corresponding solution of the GREM with the same parameterization.
Since there is a one-to-one correspondence between solutions of the GREM and solutions of the multi-$p$-spin spherical model, we expect all the quantities of interest to have the same formal dependence on the temperatures and on the universal function $q_u(y)$.
This correspondence prompt us to conjecture that the structure of the correlations
between the weights is the same in the two models, i.e. that \emph{the reference free energy of a given cluster
at temperature \( T_{1} \) is equal to the reference free energy
of the corresponding cluster at temperature \( T_{2} \).} To be precise
one should notice that the \( q(x) \) of the GREM has a qualitatively
different shape from that of the models we are considering; indeed
the solutions of the GREM does not have a plateau at \( q_{max} \)
and furthermore \( q(x)=0 \) for \( x<T/T_{c} \). This makes no difference
at all: since the solutions has the same formal expression in terms
of the function \( y_{u}(q) \), the formal expressions of quantities
like the \( P(q_{T1T2}) \) must be the same, and the difference between
the two models is only in the actual value of \( y_{u}(q) \).

According to the previous argument the reference free energies of the various clusters are fixed at all temperatures, in particular the ordering of the clusters according to them does not change; however, this does not mean that the ordering of the clusters according to their actual weights is also conserved: this is only true on average. In other words, it is not true that to the heaviest cluster at a given temperature corresponds the heaviest cluster at a lower temperature.

\section{Solutions with Different Parameterizations}

In the previous sections we saw that in order to obtain the quantities of interest we should perform sums over all the different solutions of the SP equations of the functional (\ref{FUN2}). Here we intend to discuss one of the main difficulties which prevents us from performing such sums, i.e. the existence of an infinite set of solutions corresponding to different parameterizations of the $2n\times 2n$ order parameter $\hat{Q}$.   

By applying the same technique of section 2 to the REM \cite{REM} we found the same essential features of the problem.
Here the solution of the standard isothermal problem is a very simple $q(x)$ whose value is $0$ for $x<1/\beta$ and $1$ for $x>1/\beta$; this is the simplest case of a solution that verifies the so-called Parisi-Toulouse scaling $q(x,T)=q_u(\beta x)$ \cite{PAT}, which is also found in models with full RSB like those we are considering.

As we saw in section 2 the problem is connected to the extremization of the quantity $\ln \overline{Z^{n}(T_{1})Z^{n}(T_{2})}$ with respect to the order parameter which is a \( 2n\times 2n \)
matrix \( \hat{Q}=\left( \begin{array}{cc}
Q_{1} & P\\
P^{t} & Q_{2}
\end{array}\right)  \), where \( Q_{1}, \) \( Q_{2} \) and \( P \) are \( n\times n \)
matrices. One can show that there exist solutions of the type (\ref{solq},\ref{solp}) for the REM; in particular, we have that $Q_1$,$Q_2$ and $P$ are 1RSB matrices with values $0$ for $x<1/(\beta_1+\beta_2)$ and $1$ for $x>1/(\beta_1+\beta_2)$.

In the solutions (\ref{solq},\ref{solp}) the matrices  \( Q_{1}, \) \( Q_{2} \) and \( P \) are Parisi matrices, but there exist other solutions corresponding to different parameterizations. For instance we can divide each of the three matrices \( Q_{1}, \) \( Q_{2} \) and \( P \) in four blocks of size \( n/2\times n/2 \). We have then $16$ blocks and we can parameterize each of them as a Parisi matrix. In the $n \rightarrow 0$ limit the corresponding SP equations are the same obtained by considering four different system, two of them at $T_1$ and the other two at $T_2$. The order parameter in the latter case is a \( 4n\times 4n \) matrix composed of 16 matrices of size $n \times n$ 
\begin{equation} \hat{Q}=\left( \begin{array}{cccc}
Q_{11} & Q_{12} & P_{13} &  P_{14}
\\
Q_{12} & Q_{22} & P_{23} &  P_{24}
\\
P_{13} &  P_{23} & Q_{33} & Q_{34} 
\\
P_{14} &  P_{24} & Q_{34} & Q_{44} 
\end{array}\right) 
\label{par1}
\end{equation} 
When we parameterize each of these matrices through Parisi functions $q_{11}(x)$,$q_{12}(x)\ldots$ and take the limit $n\rightarrow 0$ the corresponding equations are identical as they would be if the size of the matrices were $n/2$ rather than $n$; in other words, the solutions of the four-system problem are also solutions of the two-system problem, and they offer another way of parameterizing the order parameter \( \hat{Q}=\left( \begin{array}{cc}
Q_{1} & P\\
P^{t} & Q_{2}
\end{array}\right)  \). This argument can be extended indefinitely considering at the same time a general number of $2p$ systems, $p$ at temperature $T_1$ and $p$ at temperature $T_2$. In this way an infinite set of possible parameterizations of the $2n\times 2n$ order parameter $\hat{Q}$ in terms of Parisi matrices is obtained .  

However, not all the parameterizations give new solutions.
Going back to parameterization (\ref{par1}) we have $16$ Parisi matrices to consider; in the REM   we have a solution in which they are all equal 1RSB matrices with a breaking point $x_c=1/(2\beta_1+2\beta_2)$. It can be seen that this is nothing but a permutation of the solution in which $Q_1$,$Q_2$ and $P$ are 1RSB matrices with $x_c=1/(\beta_1+\beta_2)$ (see also the discussion in the Conclusions). Therefore the $4$-system parameterization does not add anything new in this case.
However, the 4-system parameterization describes also the following solution
\begin{eqnarray}    
q_{11}(x)=q_{22}(x)=q_{33}(x)=
q_{12}(x)=
p_{13}(x)=
p_{23}(x) & = & 
\left\{ 
\begin{array}{ccc} 
0 & \textrm{for} & x<1/(2\beta_1+\beta_2)
\\
1 & \textrm{for} & x>1/(2\beta_1+\beta_2)
\end{array} 
\right.
\nonumber
\\
q_{44}(x) & = & 
\left\{ 
\begin{array}{ccc} 
0 & \textrm{for} & x<1/\beta_2
\\
1 & \textrm{for} & x>1/\beta_2
\end{array} 
\right.
\nonumber
\\
p_{14}(x)=p_{24}(x)=q_{34}(x)
& = & 
0
\label{sol3}
\end{eqnarray}
This solution cannot be obtained through a permutation from the solution of the type (\ref{solq},\ref{solp}) and should be counted separately. Essentially, this parameterization corresponds to two systems (system $1$ and $2$) at temperature $T_1$ correlated with a system at temperature $T_2$ (system $3$), while system $4$ at $T_2$ is completely uncorrelated to the remaining three.
Accordingly, in summing over all the different solutions we must consider the infinite set of solutions whose parameterization corresponds to a generic number of $m_1$ systems at temperature $T_1$ correlated to a generic number of $m_2$ systems at temperature $T_2$, plus $m_2-m_1$ (if $m_2>m_1$) uncorrelated systems at temperature $T_1$  (in order to have the same number of system at each temperature). The scaling of the correlated components is $q(x,T_1,T_2)=q_u((m_1\beta_1+m_2\beta_2)x)$, i.e.  in the REM the correlated components are described by 1RSB function with $x_c=1/(m_1\beta_1+m_2\beta_2)$.

According to [I] any solution with a given parameterization can also be considered as solution of a certain constrained system. For instance, the blocks $Q_{11}$,$Q_{22}$,$Q_{33}$,$
Q_{12}$,$
P_{13}$,$
P_{23}$, in the solution (\ref{sol3}) correspond to a system of three real replicas which are constrained to have maximum overlap (i.e. $1$ in the REM) among themselves. When considering a constrained system the problem of summing over all solutions greatly simplifies; indeed in this  case we have only one solution plus its natural permutations and we can perform the sum in the standard way by replacing it with a sum over indices. The result is immediate, for the $P(q)$ of the constrained system we have $P(q)=\delta(q)/(2\beta_1+\beta_2)+\delta(q-1)(1-1/(2\beta_1+\beta_2))$. We define the ``free energy'' $f_\alpha$ of a state of the global system (composed of two constrained systems) as $P_\alpha=\exp [-f_\alpha]/\sum_\beta \exp [-f_\beta]$ (see footnote 1). Then we obtain that these free energies are independent random variables such that the average number of states with free energies between $f$ and $f+df$ is 
 $d\mathcal{N}(f)=\exp[f/(2\beta_1+\beta_2)]df$. This result is precisely what is obtained noticing that the states of the constrained system are triplets of identical states of the single system; therefore in the REM  their free energy is simply given by $f_\alpha^{\textrm{constrained}}=(2\beta_1+\beta_2)f_\alpha^{\textrm{single}}$ (see footnote 1) and the distribution of $f^{\textrm{single}}$ is proportional to $\exp[\beta_1 x_1 f]=\exp[\beta_2 x_2 f]=\exp[f]$. The same line of reasoning applies to the models we considered in the previous sections as well: it ensures that the results obtained by considering constrained systems (the only case in which we are able to sum over the solutions explicitly) are fully consistent with our guess on the distribution of the free energies made in section 3.

\section*{Conclusions}

We proved the existence of correlations between the equilibrium states at different temperatures of 
the spherical spin-glass models with multi-$p$ spin interactions for values of the coupling constants such that they display full RSB: there is no chaos in temperature in such models.
Furthermore, the overlaps between states at different temperatures satisfy ultrametric relations. 
Ultrametricity determines a one-to-one correspondence between the trees of states at different temperature: to any cluster of states at temperature $T_1$ at a given level of the tree corresponds a {}``twin{}" cluster of states at temperature $T_2$; the precise meaning of this correspondence has been discussed in details in the introduction and in section 2.
From a purely geometrical point of view we may say that there is only one tree of states at all temperatures, whose details are revealed lowering the temperature.
Prompted by some technical features of the problem, we conjectured that the reference free energies of two twins clusters are equal at all temperatures, exactly as in the GREM. 
Consequently, we expect that quantities like the \( P(q_{T1T2}) \) in these models are the same as the GREM. More precisely we expect them to have the same formal dependence on the temperatures and on the function \( y_{u}(q) \), whose actual values will be different in the two classes of models. 

The scenario of a tree of states which bifurcates when lowering the temperature was suggested early in spin-glass studies \cite{treecit} and was later advocated in order to explain the phenomenology of rejuvenation and memory in spin-glasses \cite{Met}. 
According to our findings, the basic premises of that phenomenological picture of off-equilibrium dynamics hold for the class of models considered here;
we believe that this further increases the need for a complete understanding of the relationship between the equilibrium energy landscape and the off-equilibrium dynamics in spin-glasses.

Within the TAP approach \cite{MPV,TAP} strong correlations between
the equilibrium states at different temperatures are readily obtained in the spherical
\( p \)-spin model with 1RSB \cite{CS,MPK,CSTAP}. 
Indeed in this model 
the angular and the self-overlap contributions to the TAP free energy
can be factored; that is if we write \( m_{i}=q^{1/2}\widehat{m_{i}} \)
, we have that the angular part \( \widehat{m_{i}} \) enters the
expression of the TAP free energy only in the form \( E_{p}=q^{p/2}\sum J_{i_{1}i_{2}\ldots i_{p}}\hat{m}_{i_{1}}\hat{m}_{i_{2}}\ldots \hat{m}_{i_{p}} \).
Accordingly, the angular components of the states, which are the extrema of the free energy, are the same at all temperatures.
To our knowledge this was the only result on the absence of chaos previously obtained for a spin-glass model.
The previous argument cannot be used in dealing with spherical spin-glass models with multi-$p$ interactions for values of the coupling constants such that they display either 1RSB or full-RSB. For instance, in the \( q^{2}-q^{4} \)
model there is an angular dependence of the form \( E=q\sum J_{ij}\hat{m}_{i}\hat{m}_{j}+q^{2}\sum J_{ijkl}\hat{m}_{i}\hat{m}_{j}\hat{m}_{k}\hat{m}_{l} \) and no factorization is possible. Indeed, when the temperature changes the self-overlap changes with it causing
the angular landscapes corresponding to the two interactions to interpenetrate. 
It would be interesting to understand how this process leads to the properties we found within the replica approach. 
Furthermore, we recall that the off-equilibrium dynamics \cite{CKBM} is connected to the TAP free-energy landscape \cite{Bir}. 
  
\begin{figure}[htb]
\begin{center}
\epsfig{file=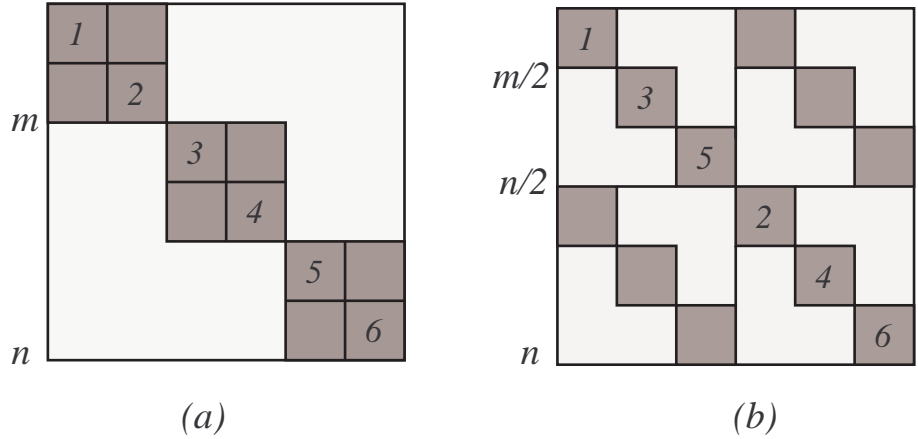}
\caption{$(a)$ The solution of a generic spin-glass problem with 1RSB. $(b)$ A permutation of the solution $(a)$ which corresponds to a parameterization of the $n\times n$ order parameter as four $n/2 \times n/2$ Parisi matrices. The generalization to full RSB is straightforward. This figure
shows that the solutions (\ref{soliso}) are obtained from a permutation of the standard Parisi solution $q_{Parisi}(x)$, in particular  
the origin of the small-$x$ scaling $q(x)=p(x)=q_{Parisi}(2x)$ is clarified.}
\label{figure0}
\end{center}\end{figure}

As discussed in [I], the solutions (\ref{solq},\ref{solp}) have the same free energy of the $P=0$ solution. This can be also derived noticing that they form a continuous line in the space of the matrices $\hat{Q}$ parameterized by the value of the continuous parameter $p_d$. The value $p_d=0$ corresponds to the standard Parisi solutions at the two temperatures. On this continuous line of solutions we have $\partial F/\partial \hat{Q}=0$ by definition; therefore the free energy is constant and equal to that of the $p_d=0$ solution.
The previous argument has deep consequences.
It provides an easy way to understand the Goldstone Theorem for disordered systems with full RSB which has been proved and discussed in \cite{KdDT}. 
This theorem connects the presence of Goldstone modes in spin-glasses with continuous RSB with the fact that the discrete permutational symmetry within the replica approach becomes a continuous symmetry when the number of RSB steps goes to infinity.
Consider the standard replica formulation of a given model: it deals with a $n \times n$ order parameter $Q_{ab}$. Instead of making the Parisi ansatz on it, we can divide it in four $n/2 \times n/2$ blocks and make the Parisi ansatz on each of these blocks separately, i.e. we assume that the two diagonal blocks are equal to an ultrametric $n/2\times n/2$ matrix $q_{ab}$ and the two off-diagonal blocks are equal to a $n/2 \times n/2$ ultrametric matrix $p_{ab}$. Then the SP equations are written in terms of two functions $q(x)$ and $p(x)$ defined on the interval $[1,n/2]$, but in the $n \rightarrow 0$ limit the equations are identical to those obtained considering $2n$ replicas with an order parameter $\hat{Q}$ parameterized by four $n \times n$ matrices, the two diagonal ones equal to $Q_{ab}$ and the two off-diagonal ones equal to $P_{ab}$. This is a special case ($T_1=T_2$) of the problem we considered in this paper. In this case \cite{FPV1} the solutions corresponding to (\ref{solq},\ref{solp}) read
\begin{eqnarray}
 q(x)=p(x)=q_{Parisi}(2x) & 0\leq x\leq \frac{1}{2}x_{Parisi}(p_{d})\nonumber
 \\
 q(x)=p(x)=p_{d} &
 \frac{1}{2}x_{Parisi}(p_{d})\leq x\leq x_{Parisi}(p_{d})\nonumber
 \\
\label{eqts}
 q(x)=q_{Parisi}(x);p(x)=p_{d}& x_{Parisi}(p_{d})\leq x\leq 1
\label{soliso}
\end{eqnarray}
Where $q_{Parisi}(x)$ is the standard Parisi solution and $p_d$ is a continuous parameter with range $[0,q_{EA}]$; every value of $p_d$ specifies a solution. As stated above, these solutions, first obtained considering $2n$ replicas, solve the standard $n$-replica problem too. 
At this point we can make the following statements:
\begin{itemize}

\item These solutions exist in every model with Parisi RSB, either discrete of continuous. 
This statement has already appeared in \cite{FN} where it is claimed that it is implied by ultrametricity.
We justify this result noticing that these solutions are {\em permutations} of the standard Parisi solution, as it is readily understood considering fig. (\ref{figure0}). 

\item When the model has full RSB the parameter $p_d$ is continuous, therefore the solutions form a continuous line of constant free energy. As a consequences on every point of the line the Hessian has zero eigenvalues.

\end{itemize}
{\bf Acknowledgements}. I thank S. Franz, G. Parisi and L. Peliti for interesting discussions. It's a pleasure to thank my family for constant help and support.

\section*{Appendix 1}
In this appendix we derive the formulas to compute a generic
function (e.g. inverse, logarithm\ldots ) of one or more Parisi
matrices. Using the standard eigenvalue technique one encounters some difficulties. 
The main problem is that in the $n \rightarrow 0$ limit the matrix ceases to be determined univocally by the set of its eigenvalues. 
For a generic Parisi matrix parameterized as $(a_d,a(x))$ the
eigenvalues are \cite{CS}
\begin{eqnarray}
\l_a (0) = & a_d-\int_0^1 a(y)dy  & \textrm{deg}:\ 1
\label{eige0}
\\
\l_a (x) = & a_d-x a(x)-\int_x^1 a(y)dy  & \textrm{deg}:\ -n\frac{dx}{x^2}
\label{eige1}
\end{eqnarray}
Two important eigenvalues are $\l_a(0)=a_d-\overline{a}$ and $\l_a(1)=a_d-a(1)$, in this context the bar means integration of the Parisi function $a(x)$ over the interval $[0,1]$.
By direct inspection one can check that the eigenvalues verify the property
$\l_{a*b}(x)=\l_a(x)\l_b(x)$, where $a*b(x)$ is the product of
Parisi algebra
\begin{eqnarray}
(a*b)_d&= &a_db_d-\int_0^1a(y)b(y)dy=a_db_d-\overline{ab}
\\
(a*b)(x)&=
&(a_d-\bar{a})b(x)+(b_d-\bar{b})a(x)-\int_0^x(a(x)-a(y))(b(x)-b(y))dy
\end{eqnarray}
It can be easily checked that the $n\ \times n$ matrix $\C$ that
projects on the vector of constant coordinates (i.e. $\C_{ab}=1\
 \forall ab$ ) has all zero eigenvalues in the limit $n\rightarrow
0$. As a consequence the set of eigenvalues ceases
to determine the matrix $A$ in a unique way.
To recover the function $(a_d,a(x))$ we need further information e.g. $a(0)$.
Using $a(0)$  and the relation $x\dot{a}(x)=-\dot{\lambda }_a(x)$ (the dot means derivative with
respect to $x$) which follows from the expressions (\ref{eige0},\ref{eige1}) we obtain the following inversion relations
\begin{equation}
a(x)=a(0)-\int_0^x \frac{dy}{y}\dot{\lambda }_a=a(0)+\frac{\lambda
_a(0)-\l _a (x)}{x}+\int_0^x \frac{dy}{y^2}(\l _a(0)-\l
_a(y))\label{a0x}
\end{equation}
\begin{equation}
a_d=\l_a(1)+a(1)=a(0)+\l
_a(0)-\int_0^1 \frac{dy}{y^2}(\l _a(y)-\l _a(0))\label{a0d}
\end{equation}
Given a function $f[a]=\sum_k f_k a^k$ we want to compute $f[A]$ for a generic ultrametric matrix A. The eigenvalues $\l_{f[A]}$ of $f[A]$ are readily obtained 
\begin{equation} 
\l_{f[A]}(x)=f[\l_a(x)]
\end{equation}
To use the formulas (\ref{a0x},\ref{a0d}) we need to know $f[A](0)$. The expression for a product is
\begin{equation}
(ab)(0)=(a_d-\bar{a})b(0)+(b_d-\bar{b})a(0)=\l _b(0)a(0)+\l
_a(0)b(0)\label{prod}
\end{equation}
As a consequence we have the following expression for the powers of $A$
\begin{equation}
(a^{n+1})(0)=\l_a(0)(a^n)(0)+\l_a(0)^n a(0) \longrightarrow (a^n)(0)=n (a_d-\overline{a})^{n-1}a(0),
\end{equation}
therefore
\begin{equation}
f[A](0)=\sum_k f_k (a^k)(0)=\sum_k f_k k (a_d-\overline{a})^{k-1}a(0)=a(0)\frac{df}{da}[a_d-\overline{a}].
\end{equation}

Summing up, the expression of a generic function $f[A]$ is
\begin{eqnarray}
f[A](x) & = &a(0)\frac{df}{da}[a_d-\bar{a}]+\int_0^xdy\
\dot{a}\frac{df}{da}[\l _a(y)]\label{fax}
\\
{1\over n }\Tr
f[A] & = & a(0)\frac{df}{da}[a_d-\bar{a}]+f[a_d-\bar{a}]-\int_0^1\frac{f[\l _a(y)]-f[a_d-\bar{a}]}{y^2}dy\label{fad}
\end{eqnarray}
The generalization to a function $g[A_1,\ldots, A_p]$ of $p$ Parisi
matrices the function is straightforward 

\begin{eqnarray}
g[A_1,\ldots ,A_p](x)& = & \sum_{i=1}^pa_i(0){\partial g\over
\partial a_i}[a_{1d}-\bar{a}_1,\ldots
,a_{pd}-\bar{a}_p]+
\\
 & + & \int_0^xdy\sum_{i=1}^p\dot{a}_i{\partial g\over
\partial a_i}[\l _{A_1},\ldots,\l
_{A_p}] \label{gx}
\end{eqnarray}

\begin{eqnarray}
{1\over n }\Tr
g[A_1,\ldots A_p] & = & \sum_{i=1}^pa_i(0){\partial g\over
\partial a_i}[a_{1d}-\bar{a}_1,\ldots,
a_{pd}-\bar{a}_p]+
g[a_{1d}-\bar{a}_1,\ldots,
a_{pd}-\bar{a}_p]+
\\
& - & \int_0^1\frac{g[\l _{A_1},\ldots,\l
_{A_p}](y) -g[a_{1d}-\bar{a}_1,\ldots,
a_{pd}-\bar{a}_p]}{y^2}dy
\end{eqnarray}

\section*{Appendix 2}
In this appendix we check that the solutions (\ref{solq},\ref{solp}) verify the
SP equation. We consider the general situation where a non zero magnetic field is present. The Free energy functional read
\begin{displaymath}
F_n=-\beta_1^2\sum_{ab}(f(Q_{1ab})+H^2Q_{1ab})-\beta_2^2\sum_{ab}(f(Q_{2ab})+H^2Q_{2ab})
\end{displaymath}
\begin{equation}
-2\beta_1\beta_2\sum_{ab}(f(P_{ab})+H^2P_{ab})-\Tr \ln \hat{Q}-n
\end{equation}
The dependence on $\hat{Q}$ can be simplified trough $\Tr \ln
\hat{Q}=\Tr \ln (Q_1Q_2-P^2)$. The SP equations then
read
\begin{equation}
\beta_1^2f'(Q_{1\
ab})+\beta_1^2H^2=-\left(\frac{1}{\hat{Q}}\right)_{1ab}=\frac{-Q_2}{Q_1Q_2-P^2}\label{spq1}
\end{equation}
\begin{equation}
\beta_1\beta_2f'(P_{ab})+\beta_1\beta_2H^2=-\left(\frac{1}{\hat{Q}}\right)_{12ab}=\frac{P}{Q_1Q_2-P^2}
\label{spp}\end{equation}
We recall the definition of the
temperature-independent function $y_u(q)$ and of its inverse $q_u(y)$
\begin{equation}
y_u(q)={f'''(q)\over 2(f''(q))^{{3\over 2}}}
\end{equation}
The self-overlap $q_{EA}(\beta)$  of the equilibrium states at a certain temperature is given by
\begin{equation}
q_{d}-q_{EA}=\frac{1}{\beta\sqrt{f''(q_{EA})}}
\end{equation}
The minimum overlap $q_H$ in presence of a magnetic field $H$ is independent of the temperature and is given by
\begin{equation}
H^2=q_Hf''(q_H)-f'(q_H)
\end{equation}
In presence of a magnetic field the generalization of the solutions (\ref{solq},\ref{solp}) is
\begin{eqnarray}
q_{s}(x) & = & \left\{ \begin{array}{ccc}
q_H & \textrm{for} & 0 \leq x\leq \frac{1}{\beta _{1}+\beta _{2}}y_{u}(q_H)\\
q_{u}((\beta _{1}+\beta _{2})x) & \textrm{for} & \frac{1}{\beta _{1}+\beta _{2}}y_{u}(q_H) \leq x\leq \frac{1}{\beta _{1}+\beta _{2}}y_{u}(p_{d})\\
p_{d} & \textrm{for} & \frac{1}{\beta _{1}+\beta _{2}}y_{u}(p_{d})\leq x\leq \frac{1}{\beta _{s}}y_{u}(p_{d})\\
q_{u}(\beta _{s}x) & \textrm{for} & \frac{1}{\beta _{s}}y_{u}(p_{d})\leq x\leq \frac{1}{\beta _{s}}y_{u}(q_{EA}(\beta_s))
\\
q_{EA}(\beta _{s}) & \textrm{for} & \frac{1}{\beta _{s}}y_{u}(q_{EA}(\beta_s))
\leq x\leq 1
\end{array}\right. 
\\
p(x) & = & \left\{ \begin{array}{ccc}
q_H & \textrm{for} & 0 \leq x\leq \frac{1}{\beta _{1}+\beta _{2}}y_{u}(q_H)\\
q_{u}((\beta _{1}+\beta _{2})x) & \textrm{for} & \frac{1}{\beta _{1}+\beta _{2}}y_{u}(q_H) \leq x\leq \frac{1}{\beta _{1}+\beta _{2}}y_{u}(p_{d})\\
p_{d} & \textrm{for} & \frac{1}{\beta _{1}+\beta _{2}}y_{u}(p_{d})\leq x\leq 1
\end{array}\right. 
\end{eqnarray}
These solutions are valid for any couple of temperatures $T_1\geq T_2$ both below the
critical temperature and for any $p_d$ between $q_H$ and $q_1(1)$
which is the self-overlap of the states of the system at the higher
temperature.
To check that they verify the SP equation
(\ref{spq1},\ref{spp}) we need to express their r.h.s. in the
Parisi form, this is readily done applying equation (\ref{gx})
\begin{eqnarray}
\left(\frac{-Q_2}{Q_1Q_2-P^2}\right)(x) & = & \frac{(q_{2d}-\bar{q}_2)^2q_1(0)+(p_d-\bar{p})^2q_2(0)-2(q_{2d}-\bar{q}_2)(p_d-\bar{p})
p(0)}{((q_{2d}-\bar{q}_2)(q_{1d}-\bar{q}_1)-(p_d-\bar{p})^2
)^2}+\nonumber
\\
 & + &\int_0^xdy\frac{\l _{q_2}^2\dot{q} _1+\l _{p}^2\dot{q} _2-2\l
_{q_2}\l _p\dot{p}}{(\l _{q_1}\l _{q_2}-\l _{p}^2)^2}
\\
\left(\frac{P}{Q_1Q_2-P^2}\right)(x) & = & -\int_0^xdy\frac{\l _{q_2}\l _p\dot{q} _1+\l _{p}\l _{q_1}\dot{q}
_2-\l _{p}^2\dot{p}-\l _{q_2}\l _{q_1}\dot{p}}{(\l _{q_1}\l
_{q_2}-\l _{p}^2)^2}
+
\nonumber
\end{eqnarray}
\begin{equation}
-\frac{(p_d-\bar{p})(q_{1d}-\bar{q}_1)q_2(0)+
(p_d-\bar{p})(q_{2d}-\bar{q}_2)q_1(0)-
(p_d-\bar{p})^2p(0)-(q_{1d}-\bar{q}_1)(q_{2d}-\bar{q}_2)p(0)}
{((q_{2d}-\bar{q}_2)(q_{1d}-\bar{q}_1)-(p_d-\bar{p})^2
)^2} 
\end{equation}
In the small-$x$ region the three functions are equal so the
previous expressions simplify in this region to
\begin{eqnarray}
\left(\frac{-Q_2}{Q_1Q_2-P^2}\right)(x) & = & 
\frac{(\l_{q_2}(q_H)-\l_{p}(q_H))^2}{(\l_{q_2}\l_{q_1}(q_H)-\l_{p}^2(q_H))^2}q_H
+\int_{q_H}^{q(x)}\frac{(\l_{q_2}(q)-\l_{p}(q))^2}
{(\l_{q_2}\l_{q_1}(q)-\l_{p}^2(q))^2}dq
\label{inte2}
\\
\left(\frac{P}{Q_1Q_2-P^2}\right)(x) & = &
\frac{(\l_{p}(q_H)-\l_{q_2}(q_H))(\l_{p}(q_H)-\l_{q_1}(q_H))}{(\l_{q_2}\l_{q_1}(q_H)-\l_{p}^2(q_H))^2}q_H+
\nonumber
\\
& +& \int_{q_H}^{q(x)}\frac{(\l_{p}(q)-\l_{q_2}(q))(\l_{p}(q)-\l_{q_1}(q))}
{(\l_{q_2}\l_{q_1}(q)-\l_{p}^2(q))^2}dq
\label{inte1}
\end{eqnarray}
The various quantities entering the previous expressions read
\begin{equation}
\begin{array}{clr}
\l_{q_1}(q)= &
\frac{1}{\beta_1+\beta_2}\left(\frac{1}{\sqrt{f''(q)}}+\frac{\beta_2/\beta_1}{\sqrt{f''(p_d)}}\right)&
\textrm{for}\ x\leq x(p_d) \nonumber
\\
\l_{q_1}(q)= & \frac{1}{\beta_1}\frac{1}{\sqrt{f''(q)}} &
\textrm{for}\ x\geq x(p_d)
\\
 \l_{p}(p)= &
\frac{1}{\beta_1+\beta_2}\left(\frac{1}{\sqrt{f''(p)}}-\frac{1}{\sqrt{f''(p_d)}}\right)&
\textrm{for}\ x\leq x(p_d) \nonumber
\\
\l_{p}(p)= & 0 & \textrm{for}\ x\geq x(p_d)
\end{array}
\end{equation}
When evaluating the quantities entering the integrals the
dependence on $p_d$ disappears so as the discontinuity at
$x(p_d)$:
\begin{equation}
\frac{(\l_{q_2}(q)-\l_{p}(q))^2}
{(\l_{q_2}\l_{q_1}(q)-\l_{p}^2(q))^2}=\beta_1^2f''(q);\ \ \ \ \
\frac{(\l_{p}(q)-\l_{q_2}(q))(\l_{p}(q)-\l_{q_1}(q))}
{(\l_{q_2}\l_{q_1}(q)-\l_{p}^2(q))^2}=\beta_1\beta_2f''(q)
\label{exin}
\end{equation}
Evaluating the integrals in (\ref{inte2},\ref{inte1}) through (\ref{exin}) we obtain
\begin{eqnarray}
\frac{-Q_2}{Q_1Q_2-P^2}(x) & = &\beta_1^2\left(f'(q_1(x))+f''(q_H)q_H-f'(q_H)\right)=\beta_1^2(f'(q_1(x))+H^2)
\\
\frac{P}{Q_1Q_2-P^2}(x) & = & \beta_1\beta_2\left(f'(p(x))+f''(q_H)q_H-f'(q_H)\right)=\beta_1\beta_2(f'(p(x))+H^2)
\end{eqnarray}
Therefore the SP equations are verified in the small-$x$ region; now for
$x\geq x(p_d)$ we have $\dot{p}=0$ and $\l_p=0$ so the equation
for $p$ is immediately verified while $q_1(x)$ and $q_2(x)$
decouple
\begin{equation}
\frac{-Q_2}{Q_1Q_2-P^2}(x)=\frac{-Q_2}{Q_1Q_2-P^2}(x(p_d))+\int_{p_d}^{q_1(x)}\frac{dq}{\l_{q_1}^2}=
\end{equation}
\begin{equation}
=\beta_1^2\left(f'(p_d)+f''(q_H)q_H-f'(q_H)+f'(q_1(x))-f'(p_d)\right)=\beta_1^2(f'(q_1(x))+H^2)
\end{equation}
We skip the explicit evaluation of the free energy; it turns out
to be the sum of the free energies at temperature $T_1$ and $T_2$, as it
should.

\end{document}